\documentclass[reprint, amsmath,amssymb,aps,]{revtex4-2}

\usepackage{graphicx}
\usepackage{dcolumn}
\usepackage{bm}
\usepackage{xcolor}
\usepackage{lipsum}  
\usepackage{lineno}
\usepackage{hyperref}
\hypersetup{
    colorlinks,
    linkcolor={red!50!black},
    citecolor={blue!50!black},
    urlcolor={blue!80!black}
}

\newcommand{\new}[1]{{\color{blue}#1}}
\newcommand{\ndbd}{\ensuremath{0\nu\beta\beta} }
 
\newcommand{\0}{\ensuremath{0\nu\beta\beta} }

\newif\ifinstructions

\instructionstrue 
 \instructionsfalse  

\begin{document}


\title{Canadian Contributions to the Search for Neutrinoless Double Beta Decay}

\author{Thomas Brunner}
  \thanks{Corresponding author: \href{mailto:thomas.brunner@mcgill.ca}{thomas.brunner@mcgill.ca}}
\affiliation{Department of Physics, McGill University, 3600 University Street, QC H3A 2T8, Canada}

\affiliation{TRIUMF, 4004 Wesbrook Mall, Vancouver, BC V6T 2A3, Canada }

\author{Jason D.~Holt}
\affiliation{TRIUMF, 4004 Wesbrook Mall, Vancouver, BC V6T 2A3, Canada }
\affiliation{ Department of Physics, McGill University, 3600 University Street, QC H3A 2T8, Canada}

\author{David McKeen}
\affiliation{TRIUMF, 4004 Wesbrook Mall, Vancouver, BC V6T 2A3, Canada}
\affiliation{Department of Physics and Astronomy, University of Victoria, Victoria, BC V8P 5C2, Canada}


\author{Alexander Wright}
\affiliation{ Institute of Particle Physics, Canada
}%
\affiliation{%
Department of Physics, Engineering Physics and Astronomy, Queen’s University, Kingston, ON K7L 3N6, Canada }

\date{\today}

\begin{abstract}
The search for neutrinoless double beta decay has internationally been recognized as the most promising approach to determine the Majorana nature of neutrinos. This hypothesized decay would, if observed, violate lepton number in weak interactions by two units, hence, prove the existence of physics beyond the Standard Model. Current experiments with sensitivity to neutrinoless double beta decay half lives of $10^{26}$ years 
did not observe such a decay and worldwide efforts are ongoing to deploy experiments with half-life sensitivities beyond $10^{28}$ years. 
Canadian groups have been involved in this search for more than four decades. This article summarizes the historical experimental efforts and describes current Canadian contributions to neutrinoless double beta decay searches and their theoretical interpretation.  


\end{abstract}

\maketitle



\maketitle
\section{Physics Motivation} 
\label{sec:motivation}

The observation of neutrino flavour oscillations~\cite{SNO:2001kpb,KamLAND:2002uet} has demonstrated that at least two species of neutrinos have nonzero mass since the oscillation probability is proportional to $\Delta m^2$, the difference of the squared masses of the neutrinos. Current oscillation data and cosmological observations have constrained neutrino masses to $\sim0.01$-$0.3~\rm eV$~\cite{Esteban:2020cvm,Lesgourgues:2012uu,Lattanzi:2017ubx}. In the standard model (SM) of particle physics each neutrino is associated with a single two-component left-chiral spinor, $\nu_L$, while the corresponding antineutrino is its right-chiral charge conjugated partner, $\bar\nu_L$. Because there are no renormalizable operators that generate their mass, neutrinos are massless in the SM. 

To give neutrinos mass, the SM must be extended with new states that have not been directly observed in nature. At low energy, there are two possible forms that neutrino mass terms can take in the Lagrangian,
\begin{equation}
{\cal L}\supset-m_D\bar\nu_R\nu_L+{\rm h.c.},
\label{eq:dirac}
\end{equation}
or
\begin{equation}
{\cal L}\supset-m_M\bar\nu_L^c\nu_L+{\rm h.c.}
\label{eq:majorana}
\end{equation}
The operator in Eq.~(\ref{eq:dirac}) involves a new low energy degree of freedom, a neutral right-chiral two component spinor $\nu_R$ and gives rise to a Dirac mass for the neutrino. The operator of Eq.~(\ref{eq:majorana}) arises from the dimension-5 Weinberg operator $(H\bar L^c)(H^\dagger L)/\Lambda$ induced by the exchange of new heavy degrees of freedom and leads to a Majorana mass. The large hierarchy in masses between the neutrinos and the charged fermions could be due an extremely small Dirac mass in Eq.~(\ref{eq:dirac}), which would be due to a Yukawa coupling to the Higgs around fives orders of magnitude smaller than the electron's. A more satisfying explanation of the smallness of the neutrino masses would be provided by Eq.~(\ref{eq:majorana}) where one would expect $m_M$ to be suppressed by the scale of the higher dimensional mass scale, $\Lambda$, in the so-called seesaw mechanism~\cite{Minkowski:1977sc,Gell-Mann:1979vob,Yanagida:1979as,Glashow:1979nm,Mohapatra:1979ia,Schechter:1980gr}.

The mass term in Eq.~(\ref{eq:dirac}) conserves a global quantum number, lepton number (provided we endow $\nu_R$ with the same lepton number as the leptons of the SM), while that in Eq.~(\ref{eq:majorana}) violates it by two units. There are good reasons to believe that lepton number is not conserved in nature. At high temperatures it is believed that the SM violates lepton number nonperturbatively through sphaleron transitions, while conserving the difference between baryon and lepton number~\cite{Kuzmin:1985mm,Arnold:1987mh,Arnold:1987zg,Arnold:1996dy,Moore:2000mx}. This opens up the possibility to explain the observed asymmetry in the universe between baryons and antibaryons through the creation of a lepton asymmetry at high temperature in a process called leptogenesis~\cite{Fukugita:1986hr,Fukugita:2002hu,Boyarsky:2009ix,Davidson:2008bu}. Additionally, there are arguments involving black hole evaporation that indicate that a quantum theory of gravity should not allow for the conservation of global quantum numbers~\cite{Zeldovich:1976vq,Banks:2010zn}. Therefore, determining whether neutrino masses conserve or violate lepton number, that is whether they are Dirac or Majorana, can shed light on some of nature's deepest mysteries. 

To answer this question requires searching for processes where lepton number could be violated. The rates for such processes are suppressed by the extremely small values of the neutrino masses. In, for instance, medium- or high-energy collider experiments such rates are proportional to $(m_M/E)^2\lesssim10^{-16}$ where $E\gtrsim100~\rm MeV$ is some relevant experimental energy scale. This suppression means that collider-based tests of whether the light neutrino masses are Majorana are likely unfeasible. (Testing whether the masses of sterile neutrinos are Majorana could potentially be done at colliders; see, e.g.~\cite{deLima:2024ohf} and references therein. In most neutrino mass models this would imply that those of the light neutrinos are Majorana as well.)

Instead, a more promising route is to make use of Avogadro's number and the enormous number of nuclei that can be contained in a sensitive detector to look for the extremely rare process of neutrinoless double beta decay ($0\nu\beta\beta$). In this process a nucleus with atomic number $A$ and charge $Z$ transitions to one with the same atomic number but two additional units of charge along with two electrons and no (anti)neutrinos,
\begin{equation}
(A,Z)\to(A,Z+2)+2e^-,
\label{eq:0nubb}
\end{equation}
violating lepton number by two units. Nuclei that can exhibit $0\nu\beta\beta$ decay can also undergo the related $2\nu\beta\beta$ decay
\begin{equation}
(A,Z)\to(A,Z+2)+2e^-+2\bar{\nu_e},
\label{eq:2nubb}
\end{equation}
where two antineutrinos are emitted, conserving lepton number. The difference between these two processes 
is that in Eq.~(\ref{eq:0nubb}) the sum of the electrons' energies is equal the $Q$-value, i.e., the mass difference between initial and final isotope, in the reaction while in Eq.~(\ref{eq:2nubb}) it forms a continuous spectrum up to $Q$.


Assuming that exchange of the light neutrinos dominates the matrix element, measuring the $0\nu\beta\beta$ rate of a particular nucleus would yield a direct measurement of the light neutrino masses in the event that they are Majorana\new{:}, 
\begin{equation}
m_{\beta\beta}^2=\bigg|\sum_{i=1,2,3}U_{ei}^2m_i\bigg|^2.
\label{eq:mbetabeta}
\end{equation}
Here, $m_{1,2,3}$ are the masses of the light neutrinos and $U_{ei}$ are elements of the first row of the Pontecorvo–Maki–Nakagawa–Sakata matrix relating neutrino flavour eigenstates to mass eigenstates. This is discussed in the following section.

\section{Nuclear Matrix Element Challenges and Canadian Contributions}

Under the assumption that \0 decay is primarily mediated by the exchange of light neutrinos, the decay rate can be related to the key beyond-Standard-Model physics quantity, the effective Majorana neutrino mass, $\langle m_{\beta\beta} \rangle$, via
\begin{equation*}
    [T^{0\nu\beta\beta}_{1/2}]^{-1}=g_A^4G^{0\nu}|M^{0\nu\beta\beta}|^2 \bigg(\frac{\langle m_{\beta\beta}\rangle}{m_e}\bigg) ^2,
\end{equation*}
where $G^{0\nu}$ denotes a well-established phase-space factor~\cite{kotila_2012} and $\mathrm{g}_{\mathrm{A}}$ is the weak axial-vector coupling constant. Finally, $M^{0\nu\beta\beta}$ is the nuclear matrix element (NME) given by
\begin{equation*}
    M^{0\nu\beta\beta} = M^{0\nu\beta\beta}_{GT}-\big(\dfrac{g_V}{g_A}\big)^2M^{0\nu\beta\beta}_{F}+M^{0\nu\beta\beta}_{T} - 2g_{\nu}^{NN}M^{0\nu\beta\beta}_{CT},
\end{equation*}
 where $M_{\mathrm{GT}}^{0\nu}$, $M_{\mathrm{F}}^{0\nu}$, and $M_{\mathrm{T}}^{0\nu}$ are the long-range Gamow-Teller, Fermi, and tensor nuclear matrix elements, respectively. 
In addition $M^{0 \nu}_{\textrm{CT}}$ represents the recently discovered short-range contact term~\cite{Ciri18contact}, where the coupling $g_{\nu\nu}$, initially unconstrained, has since been estimated to within 30\%~\cite{Ciri21contact}.


The total NME governs the rate of \0 decay from the nuclear structure perspective, and a reliable determination has been a long-standing challenge to the theory community~\cite{Engel2017}. 
Historically such calculations have been performed within the scope of various nuclear models, as discussed at length in Ref.~\cite{Agos23RMP}, but  
more recently, Canadian-led efforts have played a significant role in the advent of a first-principles, or \textit{ab initio}, approach. Many-body methods, e.g., the valence-space in-medium similarity renormalization group (VS-IMSRG)~\cite{Stroberg2017,Stro19ARNPS}, which start from only input nuclear and weak forces, have advanced to a point where all relevant isotopes are now within reach. 

A preliminary step towards reliable $0\nu\beta\beta$ calculations was achieved with the solution of the long-standing $g_A$-quenching problem in Gamow-Teller single-beta decays~\cite{Gysbers2019}. 
In this work it was shown that with a proper ab-initio treatment of both many-body correlations and electrowweak two-body currents, consistent with the initial nuclear Hamiltonian, that predicted rates from light nuclei to $^{100}$Sn generally agreed well with experimental data, with no need to consider a quenching of $g_A$.
After adequate benchmarking against quasi-exact calculations of fictitious \0  decays in light nuclei~\cite{Yao20Bench}, the NME of $^{48}$Ca was treated within three different ab initio methods~\cite{Yao20Ca,Bell21Ge,Nova21Ca}, and collective results were found to agree within estimated theoretical uncertainties. 

The push to the heavier, experimentally important, isotopes has been more complicated. 
However, with the recent advance of ab initio to the heavy regions of the chart~\cite{Miyagi2022,Hu2022}, the VS-IMSRG has provided first extensions of NME calculations to all major players in worldwide searches, $^{76}$Ge, $^{100}$Mo, $^{130}$Te, and $^{136}$Xe~\cite{Bell23TeXe}. Here it was found, particularly with consideration of the contact term mentioned above, that NME uncertainties refined extractions of $\langle m_{\beta\beta} \rangle$ by well over an order of magnitude. More recently, there has been a strong community effort towards rigorous uncertainty quantification. With a recent joint Canada-US effort, all sources of error, from i) input nuclear forces, ii) many-body physics, and iii) the \0  operator itself were estimated in $^{76}$Ge~\cite{Belley2023}, and found to be consistent with the results of Ref.~\cite{Bell23TeXe}.

Again, we note here that the above considerations assume standard mechanism of light-neutrino exchange, though 
other mechanisms for Majorana masses are possible. 
For these mechanisms, the matrix elements can differ substantially, and an ab initio analysis is currently underway for all isotopes mentioned above.

\section{Experimental Searches for Neutrinoless Double Beta Decay}
While double beta decay ($\beta\beta$) 
was first described in 1935 \cite{Mayer}, and the prospect of neutrinoless double beta decay in the case of Majorana neutrinos was introduced in 1937 \cite{Furry}, the extremely long half lives meant that significant experimental advances were required to probe these processes at a meaningful level. Indeed, the Standard-Model allowed two neutrino double beta decay was not detected until 1987 \cite{ElliottMoe}, and the search for neutrinoless double beta decay continues very actively today. 

\begin{figure}[htb]
    \centering
    \includegraphics[width = 3.4in]{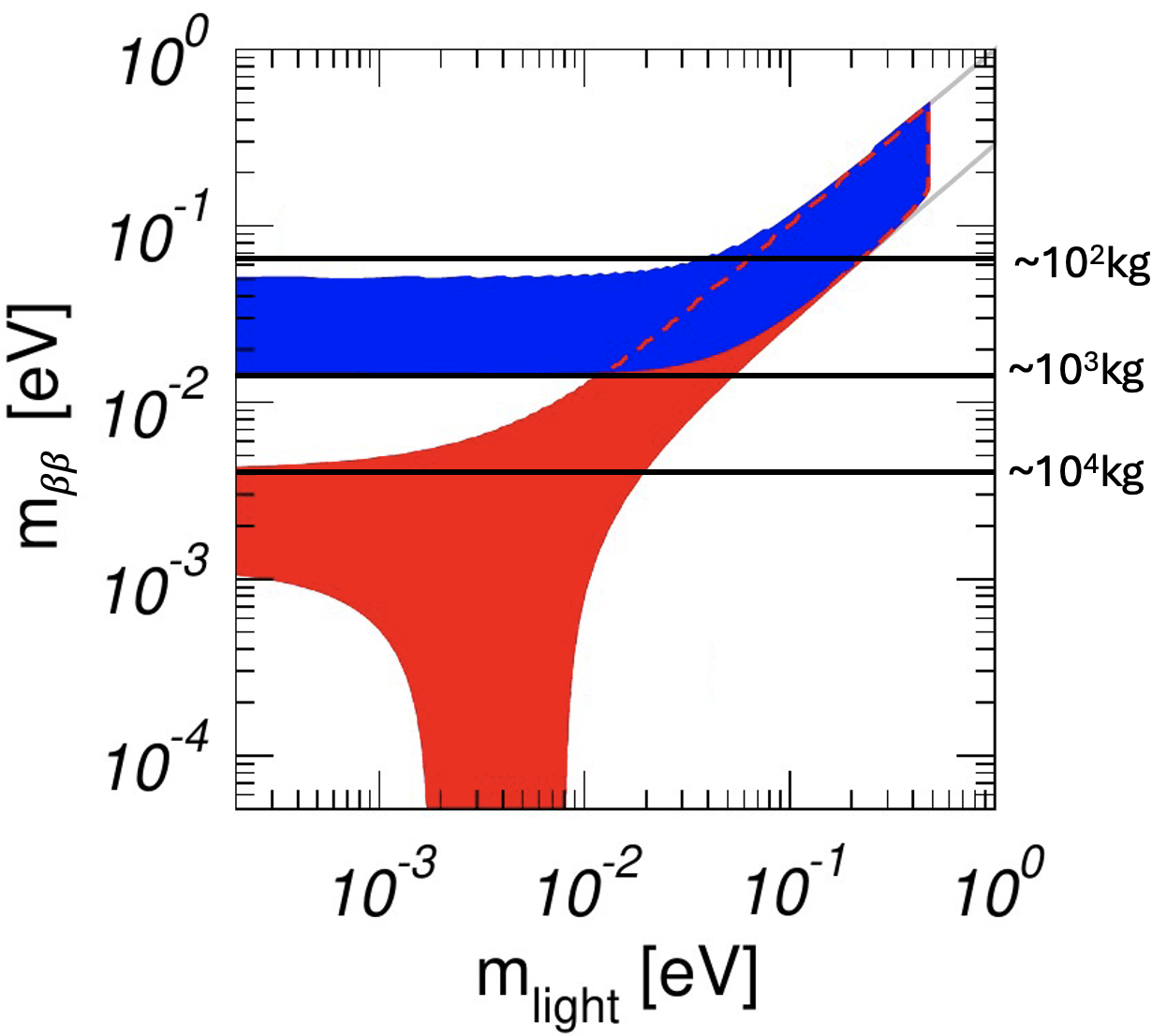}
    \caption{The neutrinoless double beta decay parameter space allowed by measured neutrino mixing parameters for normal (red) and inverted (blue) neutrino mass hierarchies. Here $m_{\beta\beta}$ is the effective Majorana neutrino mass, as introduced in Eq. \ref{eq:mbetabeta}, while m$_\text{light}$ is the mass of the lightest neutrino mass eigenstate. Horizontal lines indicate the approximate mass of $\beta\beta$-decaying isotope required for experiments to reach different levels of sensitivity. Adapted from NuFit-6.0 \cite{NuFit}, used under  \href{http://creativecommons.org/licenses/by/4.0/}{CC BY 4.0} license.}
    \label{fig:lobster}
\end{figure}

Assuming that \0 is primarily mediated by the exchange of light neutrinos, $m_{\beta\beta}$ can be parameterized by $m_0$, the lightest neutrino mass eigenstate. In this case, the 
phase space available for \0
can be 
constrained via the neutrino mass splittings and mixing parameters measured by neutrino oscillation experiments. As shown in Figure \ref{fig:lobster}, 100\,kg scale experiments, such as those currently underway, 
begin to probe the top of the inverted hierarchy, while the next generation of tonne-scale experiments should probe the majority of the inverted hierarchy parameter space and are sensitive to the normal hierarchy for cases where the mass states are close to degenerate.

Neutrinoless double beta decay is typically accessed experimentally through the precise measurement of the sum of the energy carried by the two decay electrons. As described in Section \ref{sec:motivation}, in two neutrino double beta decay the neutrinos carry some energy resulting in a continuous electron energy spectrum, while in neutrinoless double beta decay the electrons carry the full decay energy, referred to as $\textrm{Q}_{\beta\beta}$ or Q. The experimental signature of \0
would thus be a small mono-energetic peak at the endpoint of the comparatively 
large two neutrino double beta decay continuum. Experimental sensitivity to neutrinoless double beta is achieved through three main factors: maximizing the mass of target isotope in the experiment to increase the potential signal rate; achieving the best possible energy resolution to facilitate discrimination of the signal from the two neutrino double beta decay spectrum and other backgrounds; and reducing and suppressing the rate of background events in the signal region as much as possible. Many different experiments have searched for neutrinoless double beta decay using different isotopes and experimental approaches to attempt to optimize these parameters \cite{annurevDBD}.


\section{Canadian Involvement}
Canadian groups have been involved in $0\nu\beta\beta$ searches since the 1980's. Over the last two decades, Canadian efforts mainly focused on $0\nu\beta\beta$ searches in the isotope $^{136}$Xe with the EXO-200 and nEXO experiments and in $^{130}$Te with SNO+. 
Canadian researchers are also involved in searches in $^{76}$Ge with the experiments {\sc Majorana} and LEGEND-1000, and they work with CUPID collaborators on detector R\&D for advanced bolometer experiments, such as future upgrades to the CUPID experiment.

Canadian experimental $0\nu\beta\beta$ efforts  are currently primarily supported by the Canada Foundation for Innovation and provincial partners for capital investments, and the Natural Sciences and Engineering Research Council of Canada for personnel and operating costs. SNOLAB, TRIUMF, and the Arthur B. McDonald Canadian Astroparticle Physics Research Institute provide important additional support.


\subsection{Early Canadian Efforts}

\begin{figure}[h]
\includegraphics[width = 3in]{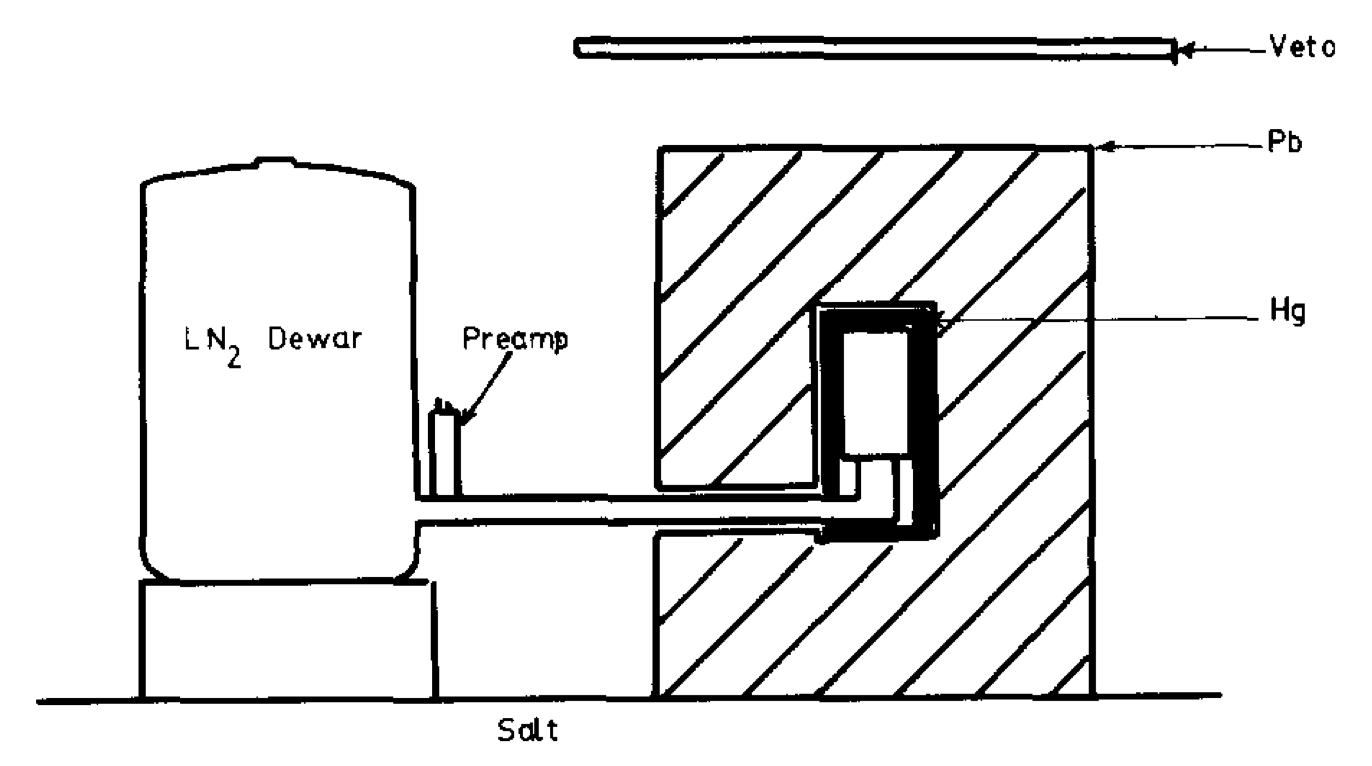}%
\caption{\label{fig:Simpson}The configuration of the Windsor Salt Mine germanium experiment. Reprinted from \cite{SimpsonNIM}, with permission from Elsevier.}
\end{figure}

The first neutrinoless double beta decay search in Canada came in the early 1980's, with the experiment of Simpson, Jagam, Campbell, Malm, and Robertson \cite{Simpson}. A custom-developed high-purity germanium detector was used which, at $\sim$1\,kg, was the largest yet employed in a double-beta decay experiment. The detector was operated in a salt mine near Windsor, Ontario, which was selected due to its very low level of radiogenic backgrounds relative to hard-rock mines. The detector was shielded with a novel lead and mercury shield (see Figure \ref{fig:Simpson}). The experiment yielded a then-competitive  upper limit on the neutrinoless double beta decay half life of $3.2\times10^{22}$ years (68\% Confidence Level (C.L.)), which corresponds to an effective Majorana neutrino mass of $\sim$18\,eV. A 3\,kg array of well shielded low background germanium detectors was subsequently developed \cite{Simpson2}, but it was mainly used for gamma-ray spectroscopy, including materials screening for the Sudbury Neutrino Observatory (SNO) experiment \cite{Simpson3}.

At around the same time as Canadian $0\nu\beta\beta$ 
searches were starting, Canadians were involved in calculations of the neutrinoless double beta decay matrix elements, and exchange reaction measurements to inform those calculations were undertaken at TRIUMF.

From the late 1980's, the development and execution of the SNO experiment became the major focus of the Canadian particle astrophysics community, and experimental work on double-beta decay searches subsided. As the SNO experiment matured in the mid 2000's, Canadian groups re-engaged in the neutrinoless double beta decay challenge, becoming involved in a number of 
experimental efforts: SNO+, {\sc Majorana}, and EXO-200, and next-generation experiments LEGEND-1000, CUPID, and nEXO. 

\subsection{SNO+}
\begin{figure}[h]
\includegraphics[width = 3in]{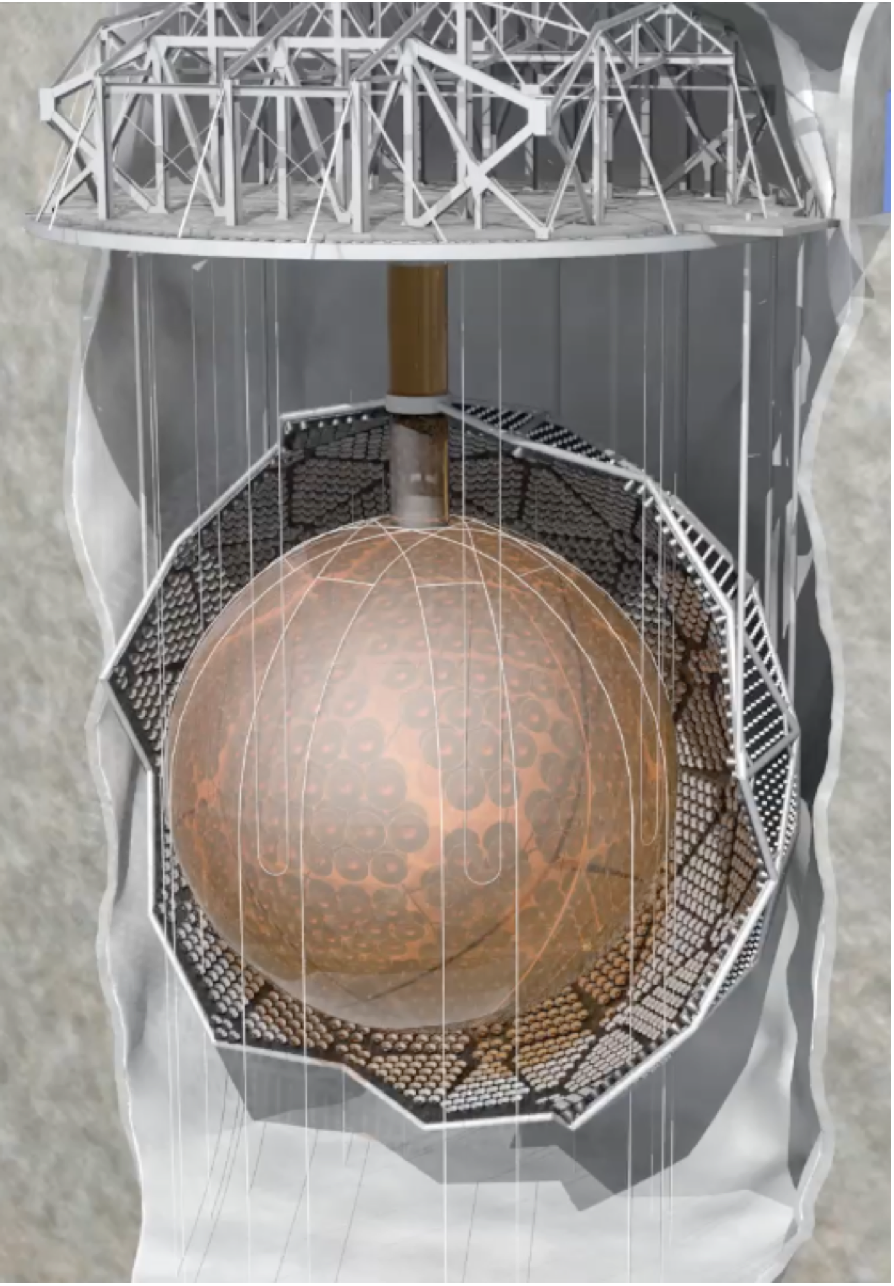}%
\caption{\label{fig:SNO+}A cutaway illustration of the SNO+ detector. 
From \cite{SNO+Detector}. $\copyright$ IOP Publishing. Reproduced with permission. All rights reserved.
}
\end{figure}

SNO+ is a Canadian-led experiment that will search for $0\nu\beta\beta$ 
by chemically dissolving a tellurium compound in a large liquid scintillator detector. The concept for SNO+ began to develop in about 2004 as the SNO experiment matured \cite{ChenFirstSNO+}. A sub-set of the SNO collaboration, led by a group at Queen's University, began discussions on the possible re-use of the SNO infrastructure for a liquid scintillator experiment. The large size and greater depth of SNO+ relative to other liquid scintillator experiments gave the experiment the potential to make a leading measurement of the low energy solar neutrinos, and reactor-antineutrino, geo-neutrino, and supernova neutrino measurements would also be possible \cite{SNO+Physics}. An artistic impression of the SNO+ detector is shown in Figure \ref{fig:SNO+}.

However, the main science target for SNO+ is the search for neutrinoless double beta decay. The SNO+ approach was based on a proposal by Ragahavan \cite{Raghavan} that a sensitive neutrinoless double beta decay search could be executed by loading isotope in a large liquid scintillator experiment. Detailed simulations of the expected performance of SNO+ confirmed that competitive sensitivity could be achieved through the isotope loading approach. The very large mass and low background achievable in a large liquid scintillator compensated for the most modest energy resolution \cite{wright}. 

Development work was carried out to investigate techniques to load $\beta\beta$ decaying 
isotopes into the SNO+ scintillator. Early loading studies included both chemical techniques and the suspension of nanoparticles in the scintillator. The latter showed the intriguing property that some atomic absorption lines appeared to be suppressed in the nanoparticle form; however, given the novel nature of nanoparticle technology at the time, conventional loading was pursued. 

Scintillator loading techniques were studied for different elements. Neodymium was an early focus, in large part due to its relatively high neutrinoless double beta decay end point energy. A loading technique based on the carboxylate technique for indium loading was developed for LENS \cite{LENS} and tested, along with a method to purify the neodymium using a barium co-precipitation approach. The loading level was limited to about 0.1\% in SNO+ by the neodymium atomic absorption, meaning that isotopic enrichment was required to achieve competitive sensitivity. Neodymium enrichment is possible through laser isotope separation \cite{AVLIS}, and the use of a large-scale AVLIS facility in France was explored but did not proceed. 

Tellurium was ultimately selected for the SNO+ $0\nu\beta\beta$ 
search due to its good optical transparency and the high natural abundance ($\sim$35\%) of the $\beta\beta$-decaying isotope $^{130}$Te, which makes enrichment unnecessary, 
keeping the experiment cost 
relatively modest. Indeed, it has been suggested that tellurium-based detectors might provide a practical path to neutrinoless double beta decay searches with sensitivity to effective neutrino masses at the level of the non-degenerate normal neutrino mass hierarchy \cite{billerNH}. A technique to purify the tellurium to the necessary levels via pH selective recrystallization was developed \cite{SNO+TePurification}, along with a technique to load tellurium into the SNO+ scintillator at concentrations up to several percent by weight (which would correspond to 5-10 tonnes of $^{130}$Te isotope in SNO+) via the production of a tellurium-organic complex soluble in the liquid scintillator \cite{SNO+TeLoading}. 

In order for SNO+ to be realized, a novel acrylic-compatible liquid scintillator based on linear alkylbenzene was developed \cite{SNO+Scintillator}, a scintillator purification and handling system was designed and installed \cite{SNO+Detector}, and a number of other hardware and electronics upgrades were carried out \cite{SNO+Detector, SNO+HoldDown}. Process systems capable of performing the tellurium purification and loading processes at the 200\,kg batch size have been developed and installed underground at SNOLAB, and a first ``test batch'' of tellurium was processed in early 2024.

SNO+ operations are divided into three phases. Initially the detector was filled with ultrapure water to protect the apparatus and facilitate scintillator loading. The detector was operated in this state as a water Cherenkov detector between 2017 and 2019, enabling low-background measurement of the $^8$B solar neutrino flux \cite{SNO+WaterSolar}; world-leading limits to be set on several invisible nucleon decay modes \cite{SNO+ND}; and the first detection of reactor anti-neutrinos in water \cite{SNO+WaterAntinu}, enabled by the improved SNO+ data acquisition which allowed neutron captures to be efficiently detected in water for the first time \cite{SNO+WaterNeutrons}. The detector was then filled with liquid scintillator. Solar neutrino and antineutrino analyses based on this ``pure scintillator'' fill are underway, and a first demonstration of event-by-event direction reconstruction in a large liquid scintillator detector has already been published \cite{SNO+ScintDirection}.

The third phase of SNO+ will be the neutrinoless double beta decay search phase with tellurium deployed in the detector. The SNO+ collaboration currently has in underground storage enough tellurium to load SNO+ with 0.5\% tellurium, which on its own is expected to enable a neutrinoless double beta decay search with half-life sensitivity of $2 \times 10^{26}$\,yr (90\% C.L.), and effective Majorana neutrino masses in the $30 - 70$\,meV range. Increasing the loading to the few percent level is expected to increase the sensitivity to $1-2 \times 10^{27}$\,yr (90\% C.L.) and $10 - 30$\,meV. SNO+ tellurium deployment is anticipated to begin during 2025.

\subsection{{\sc Majorana} and LEGEND-1000}

\begin{figure}[h]
\includegraphics[width = 3in]{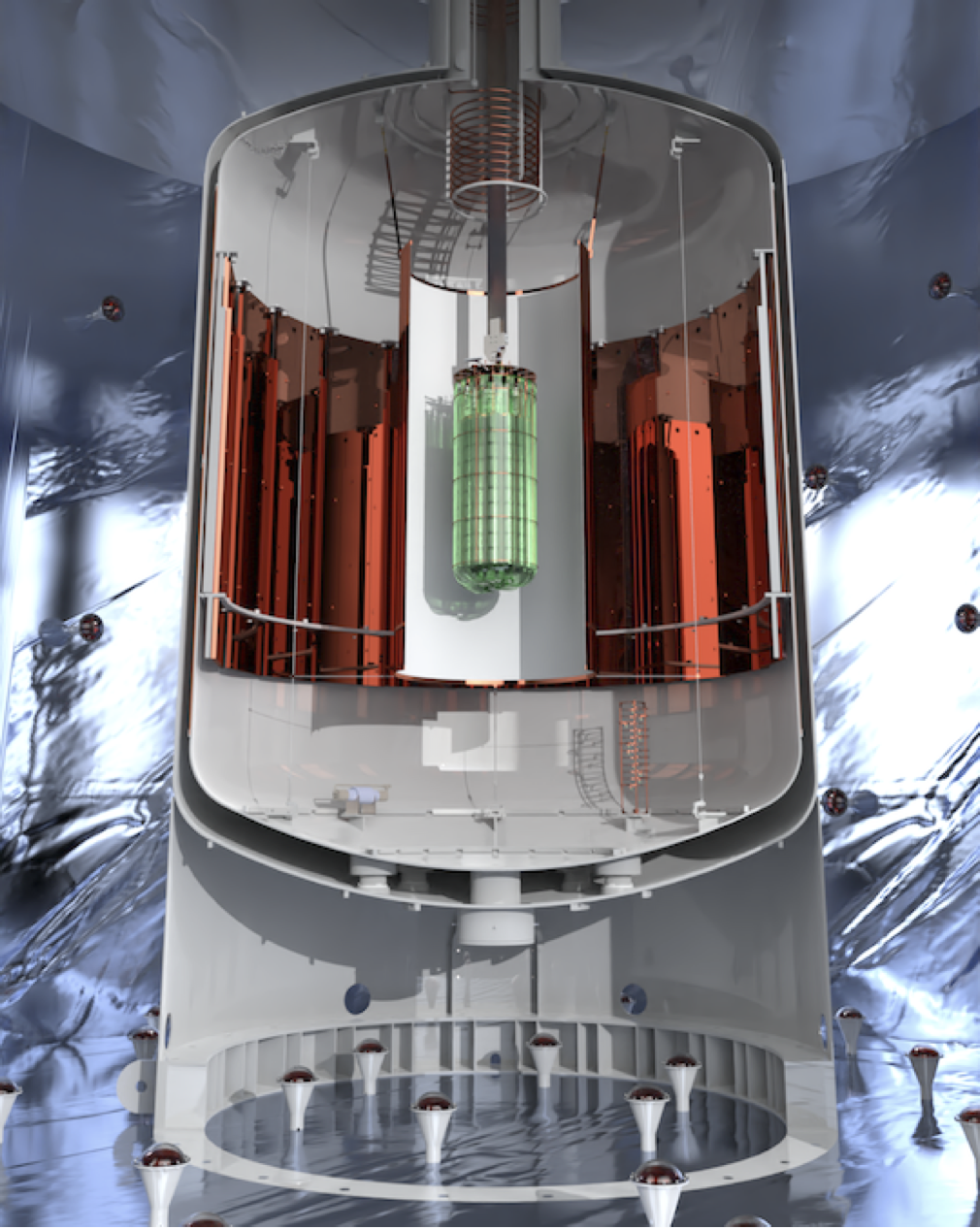}%
\caption{\label{fig:LEGEND200}A cutaway rendering of LEGEND-200 in the GERGA cryostat at LNGS. The Ge crystals are installed in ``towers'' suspended in liquid argon inside a curtain of wavelength-shifting fibers for background identification and suppression (green cylinder). Figure reproduced from \cite{LEGEND1000}, used under  \href{http://creativecommons.org/licenses/by/4.0/}{CC BY 4.0} license.}
\end{figure}

LEGEND-1000 is a proposed tonne scale neutrinoless double beta decay experiment based on enriched germanium-76 \cite{LEGEND1000}. The LEGEND collaboration was formed through the merger of the {\sc Majorana} \cite{MJD} and GERDA \cite{GERDA} collaborations, plus others. The LEGEND collaboration is currently operating LEGEND-200, an optimized re-deployment of the enriched germanium from the two preceding experiments, at the Gran Sasso National Laboratory (LNGS) in Italy. LEGEND-200 is show in Figure \ref{fig:LEGEND200}.

Canadian involvement in these efforts again began with SNO collaborators at Queen's University, who in the early 2000's joined  {\sc Majorana}. The Canadians supported the then-proposed merger of {\sc Majorana} and GERDA, and a Canadian postdoctoral researcher was an early developer on MaGe \cite{Mage}, a shared Monte Carlo simulation platform for {\sc Majorana} and GERDA - one of the first shared efforts between the two groups. These early Canadian efforts on {\sc Majorana} were not funded, however, and did not continue. 

More recently, Canadians have returned to the $^{76}$Ge efforts and currently, there are several 
Canadians involved with the LEGEND-1000 effort. Canadian contributions include germanium technology development, contributions to process and cooling systems for argon, background control and material assay, and analysis contributions including pulse de-noising and multiple scattering discrimination.


\subsection{EXO-200 and nEXO}
%
Canadian researchers are involved in the search for \ndbd in the isotope $^{136}$Xe using liquid-xenon (LXe) time-projection chamber (TPC) technology with the experiments EXO-200 and nEXO. The use of a such a TPC to search for \ndbd  was pioneered by the Enriched Xenon Observatory (EXO) collaboration with the EXO-200 experiment. Based on its success, the next-generation double-beta decay experiment nEXO is being developed to be preferentially sited at the SNOLAB Cryopit. 

In this approach, the LXe 
acts simultaneously as the decay medium, in the form of the $\beta\beta$-decaying isotope $^{136}$Xe, and the detection medium. Energy deposited within the LXe volume creates ionization electrons and scintillation light. A constant electric field is applied to the LXe volume to drift ionization electrons to a segmented anode where they are detected. Segmenting the anode allows for the reconstruction of the ionization-electron distribution projected onto the anode plane. Photosensors within the TPC detect the scintillation light. From the time difference between prompt scintillation light and ionization-electron detection, the lateral position within the detector is determined. While the anti-correlation between ionization and scintillation channels in liquid xenon limits the energy resolution of each channel individually, a simultaneous measurement of both channels significantly improves the energy resolution \cite{EXO-200:2003bso}. It further allows the suppression of alpha backgrounds, which, due to a high ionization density create more scintillation light in LXe than electron interactions (see e.g. \cite{EXO-200:2015ura}). The localization of each energy deposit within the TPC volume allows the determination of multiplicity of each event, i.e., events where all energy is deposited in one location are referred to as single-site events with multiplicity 1, events with multiplicity larger than 1 are referred to as multi-site events 
where energy is deposited at multiple locations. Double-beta decay events at the $Q_{\beta\beta}$ value ($Q_{\beta\beta}\left(\textnormal{$^{136}$Xe}\right) = 2457.83(37)$\, keV \cite{Redshaw:2007un}) typically have a multiplicity of 1 while $\gamma$-rays at this energy typically are of multiplicity $>1$, due to Compton scattering throughout the detector. This provides an effective approach to separate $\beta\beta$ like signals from background events which are dominated by $\gamma$ photons.

\subsubsection{The EXO-200 experiment}
EXO started as an R\&D effort towards a double-beta decay experiment using the isotope $^{136}$Xe in 2001 with the first Canadian groups, Carleton University and Laurentian University, joining in 2004. The EXO-200 detector was constructed at the Waste Isolation Pilot Plant (WIPP) in New Mexico, USA, from 2007 to 2010 with major contributions from the Canadian groups, for example the calibration system and radioassaying of detector materials. During EXO-200's operational phase from 2011 to 2018, two additional Canadian groups joined the collaboration, TRIUMF in 2013, and McGill University in 2015. Canadian researchers have taken leading roles in the scientific exploitation of EXO-200. For most of the analyses, one of the two analysis coordinators leading a particular analysis was from Carleton or Laurentian University. After joining EXO-200, a McGill researcher oversaw and coordinated the detector's operation as member of the EXO-200 management team. 
\begin{figure}[ht]
\includegraphics[width = 3in]{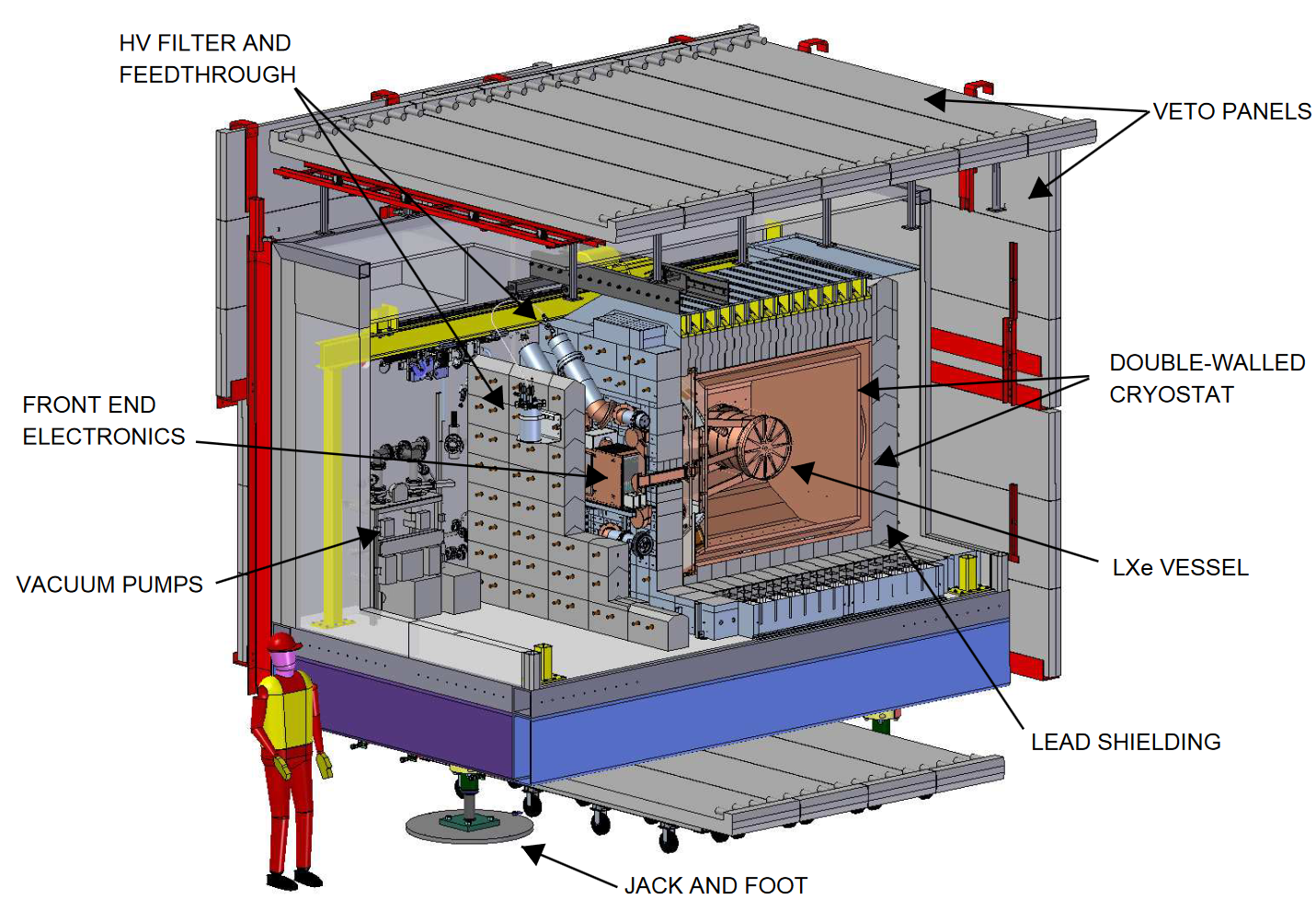} 
\caption{\label{fig:exo200}A cutaway illustration of the EXO-200 detector at the Waste Isolation Pilot Plant in NM, USA. Figure reproduced from \cite{Auger:2012gs}. $\copyright$ IOP Publishing. Reproduced with permission. All rights reserved.
}
\end{figure}

EXO-200 used an active mass of $\sim110$\,kg of LXe enriched to 80.6\% in the $\beta\beta$-decaying isotope $^{136}$Xe \cite{Auger:2012gs,EXO-200:2013xfn}. The detector consisted of two TPC halves that shared a common, optically transparent cathode. Copper rings created a homogeneous electric field to drift ionization electrons onto an anode plane consisting of two wire planes oriented at an angle of 60$^{\circ}$. One wire plane read the induction signal of drifting electrons, while the second wire plane collected the charges, allowing for a 2D reconstruction of the ionization cloud. Teflon sheets were placed in front of the field shaping rings to reflect the vacuum-ultraviolet scintillation light (175\, nm \cite{FUJII2015293}) onto large-area avalanche photo diodes. The xenon, charge-readout wire planes, and photosensors are contained within a low-radioactive copper LXe vessel. Xenon is kept liquid by submerging the TPC in the cryogenic fluid HFE-7000 inside a vacuum-insulated copper cryostat. The latter is located inside a low-radioactive lead shielding inside a cleanroom. The clean room is surrounded on five sides by muon-veto scintillator panels. A sectioned view of an engineering rendering of EXO-200 is shown in Figure\,\ref{fig:exo200}. For more details on the EXO-200 detector see \cite{Auger:2012gs,EXO-200:2021srn}. 

EXO-200 discovered the $2\nu\beta\beta$ mode in $^{136}$Xe \cite{EXO-200:2011xzf} and measured its rate with improved precision \cite{EXO-200:2013xfn}. EXO-200 published four searches for $0\nu\beta\beta$ \cite{EXO-200:2012pdt,EXO-200:2014ofj,EXO:2017poz,EXO-200:2019rkq} with a final sensitivity of $5.0\times10^{25}$ yr and a 90\% C.L. limit on the half life of this decay of $T^{0\nu}_{1/2}>3.5\times 10^{25}$ yr with 234.1 kg$\times$yr exposure \cite{EXO-200:2019rkq}. In addition, searches for exotic Majoron-decays \cite{Kharusi:2021jez}, searches for MeV dark matter recoils \cite{EXO-200:2022adi}, and $\beta\beta$-decay searches to excited states \cite{EXO-200:2023pdl} were performed using the EXO-200 dataset. 

EXO-200 demonstrated the power of using a LXe TPC in the search for $0\nu\beta\beta$ decay through energy reconstruction, event multiplicity, and event distribution in the TPC volume, i.e., the distance of an event from the nearest detector component. The detector achieved an energy resolution  of $\sigma/E = 1.15\%$ at $Q_{\beta\beta}$  \cite{EXO-200:2019rkq} and it demonstrated the validity to use radioassay results taken before and during construction to estimate the detector's sensitivity \cite{Albert:2015nta}. The successor experiment nEXO builds on the experience gained with, and the success of, the EXO-200 detector.
\subsubsection{The nEXO experiment}
\begin{figure}[ht]
\includegraphics[width = 3in]{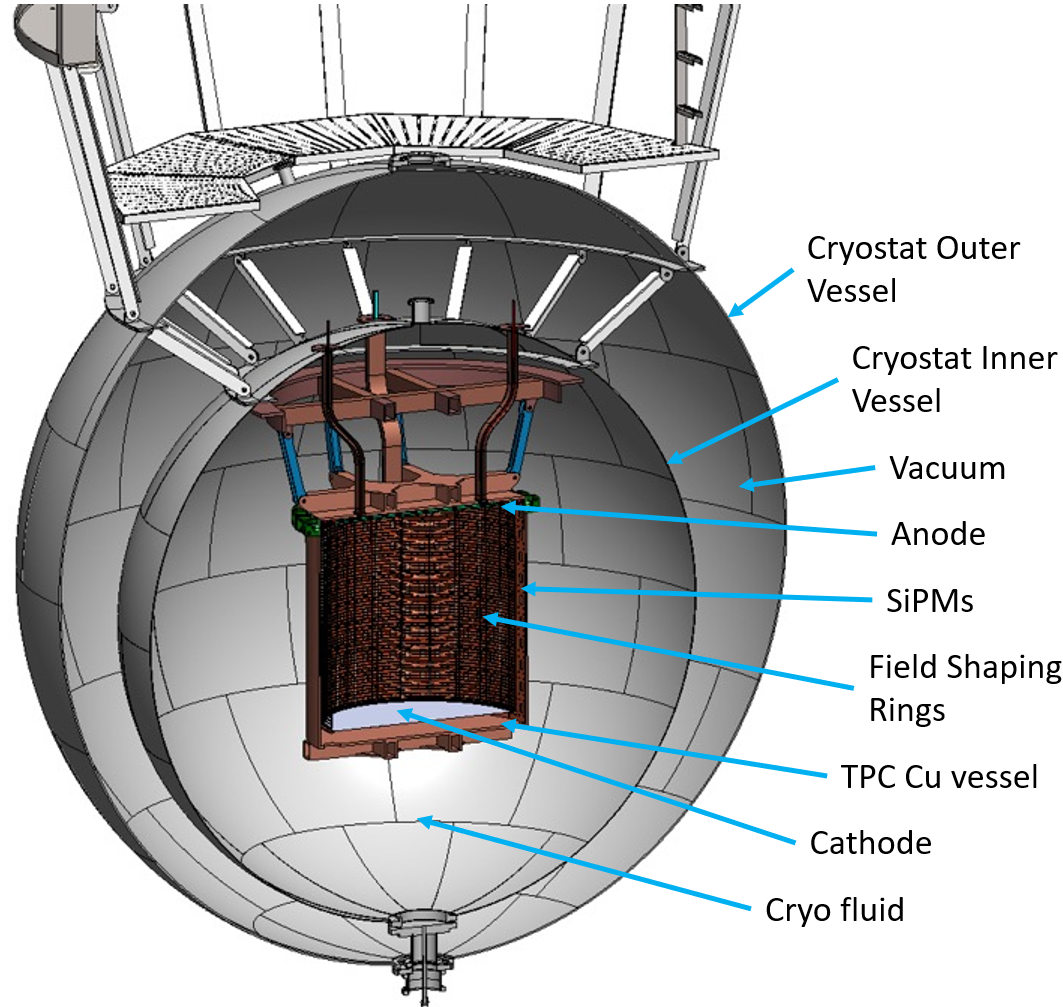}%
\caption{\label{fig:tpc}A cutaway illustration of the nEXO TPC inside the vacuum cryostat. See text for details.}
\end{figure}

\begin{figure}[ht]
\includegraphics[width = 3in]{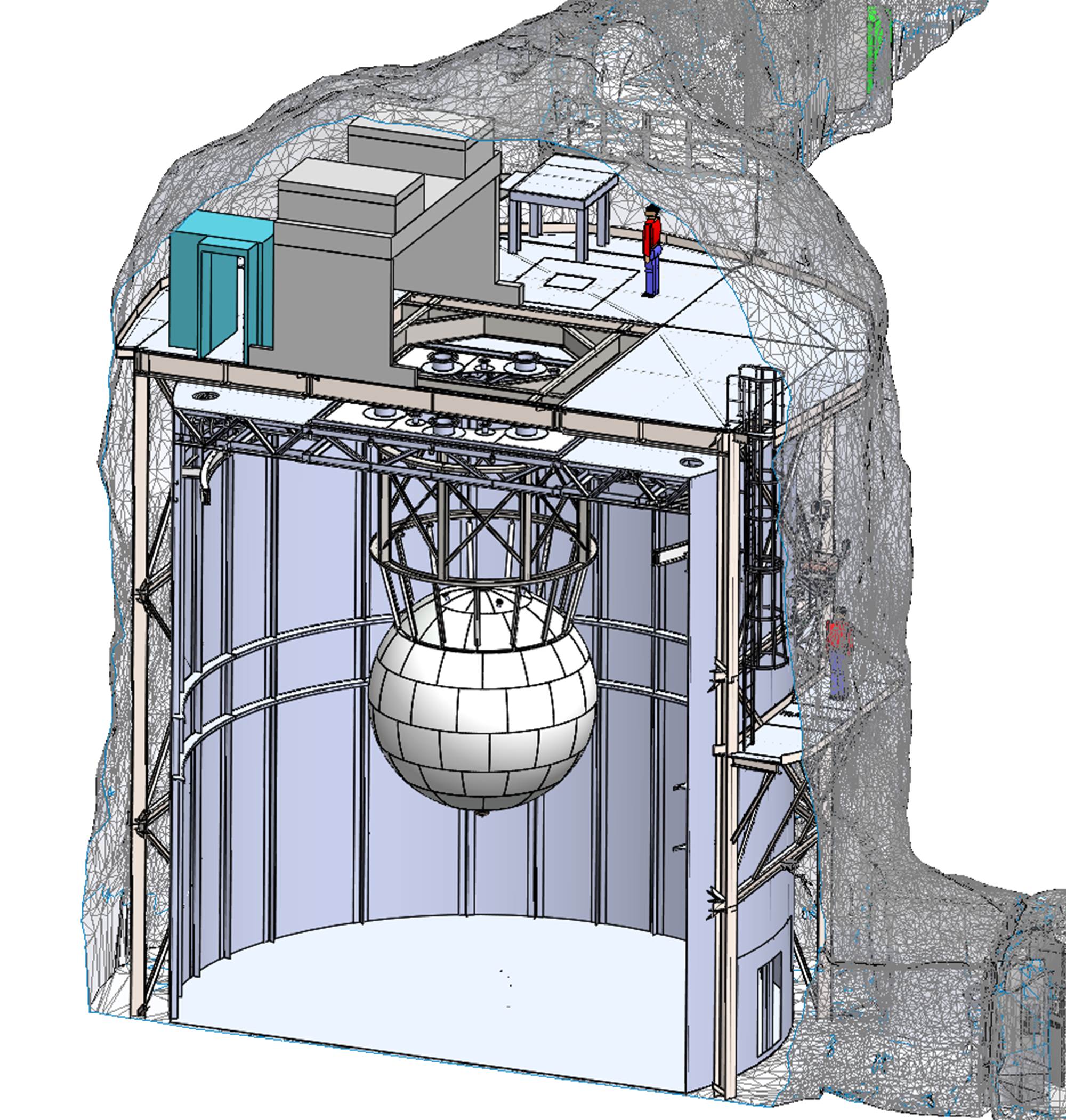}%
\caption{\label{fig:nEXO}A cutaway illustration of the nEXO detector at the SNOLAB Cryopit. The TPC is located inside a vacuum insulated cryostat that is suspended from a free-standing platform. It is located at the centre of a water tank With a diameter of 12.3\,m and a height of 12.8\,m. Figurines are shown for size comparison.}
\end{figure}

nEXO is a next-generation $0\nu\beta\beta$ decay experiment that is being developed to reach a sensitivity beyond $10^{28}$\,yr. It will deploy 5 tonnes of liquid xenon, enriched to 90\% in the isotope $^{136}$Xe, in a monolithic single-phase TPC. A sectioned view of an engineering rendering of the TPC and cryostat is shown in Figure\,\ref{fig:tpc}. Charge tiles \cite{nEXO:2017pvm,nEXO:2019nye} located at the top of the drift volume will record the ionization-electron signal while silicon photomultipliers (SiPMs) will record the scintillation light. Both signals will be digitized by cold electronics inside the TPC. The SiPMs will be mounted  inside the LXe between the TPC wall and the field shaping electrodes, facing inwards. SiPMs have been identified as the technology of choice due to their low intrinsic radioactivity, high gain, single-photon sensitivity, and low operating voltage. SiPMs have been identified that meet nEXO's requirement with respect to photodetection efficiency ($>15\%$) \cite{nEXO:2018cev,Gallina:2019fxt,Gallina:2022zjs}. This is required for nEXO to reach its anticipated energy resolution of better than 1\% at the $Q_{\beta\beta}$ value. The xenon will be kept liquid at $\sim168$\,K by submerging the TPC in a cryogenic fluid, similar to EXO-200, in a vacuum insulated cryostat. This thermal bath further acts as shielding against radioactive backgrounds. The cryostat with TPC is suspended from a free standing platform and positioned at the center of a large water tank. This tank, 12.3\,m in diameter and 12.8\,m in height, will be filled with ultra-pure water to shield the inner detector from environmental backgrounds from the experiment's surroundings. The water tank will be instrumented with photomultiplier tubes to detect Cherenkov light of passing cosmogenic radiation, primarily muons. An engineering rendering of nEXO at the SNOLAB Cryopit, the location preferred by the collaboration, is shown in Figure\,\ref{fig:nEXO}. nEXO is described in detail in \cite{nEXO:2018ylp}.  

The Canadian team on nEXO consists of groups at Carleton University, the University of British Columbia, Laurentian University, McGill University, McMaster University, Queen's University, Universit\'e de Sherbrooke, SNOLAB, TRIUMF, and the University of Windsor. The team is responsible for two of nEXO's ten detector systems: the facility infrastructure at SNOLAB and the outer detector water shield and muon veto. In addition, the TRIUMF group leads R\&D efforts towards the photodetector system with participation of Carleton, McGill, and Sherbrooke. Laurentian University, SNOLAB, and the University of Windsor are performing radioassay and Rn-emanation measurements of detector components. Queen's University is pursuing the development of electroforming infrastructure that is required to grow low-radioactive copper for nEXO's TPC vessel and associated components. 

nEXO is projected to reach a sensitivity of $1.35\times10^{28}$ yr at the 90\% C.L. after ten years of data taking with a $3\sigma$ discovery potential of $0.74\times10^{28}$ yr \cite{nEXO:2021ujk}. The physics reach of nEXO in $\langle m_{\beta\beta}\rangle$ parameter space ranges from 4.7 to 20.3 meV. 

\subsubsection{Ba-tagging technology as upgrade path to nEXO}
The $^{136}$Xe search for \ndbd  using TPC technology, either in gaseous or liquid xenon, opens the possibility for ``Ba-tagging,'' that is, the identification of the $^{136}$Xe $\beta\beta$ decay daughter isotope Ba-136 following a potential \ndbd  event \cite{Moe:1991ik}. An effective Ba-tagging approach would allow a measurement free of any backgrounds, except those from $2\nu\beta\beta$ decays. In a nEXO-type detector this would result in an increase in sensitivity of 2-3 \cite{nEXO:2021ujk}. Initially, researchers at Carleton focused on developing a Ba-tagging technique for a gaseous Xe TPC \cite{Sinclair:2011zz}. With liquid Xe TPC technology being selected for nEXO, and TRIUMF and McGill groups joining the collaboration, efforts shifted toward developing a Ba-tagging technique as an upgrade path for nEXO. This technique will employ a capillary to extract from the TPC a specific, small volume of liquid xenon surrounding the location of a potential \ndbd  event and subsequently probing it for the presence of a Ba-136 daughter ion. In detail, once outside the TPC, xenon will undergo a phase transition from liquid to gas and a radio-frequency ion funnel will subsequently extract the Ba ion from neutral Xe gas \cite{Brunner:2014sfa}. Quadrupole ion guides and ion traps will mass-select, cool, and trap the ion \cite{nEXOBa-tagginggroup:2023kmq} for laser fluorescence spectroscopy \cite{Green:2007rc} and time-of-flight mass-to-charge identification \cite{Murray:2019snw}. An accelerator-driven Ba-ion source is being developed at TRIUMF to characterize and quantize the efficiency of this and other Ba-tagging techniques that are being pursued. A comprehensive description of the Canadian Ba-tagging approach is provided in \cite{Ba-taggingCanada}, a summary of global Ba-tagging efforts is provided in \cite{Anker:2024xfz}.

\section{CUPID}
\begin{figure}[ht]
\includegraphics[width = 3in]{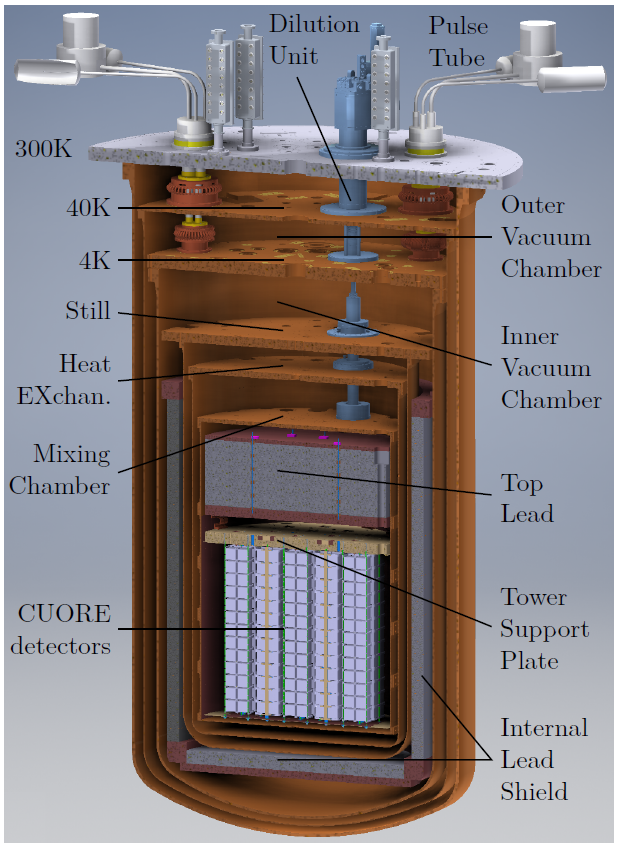}%
\caption{\label{fig:cuore}Rendering of the CUORE cryostat with different thermal stages, vacuum chambers, cooling elements, lead shields and the detector itself. The CUPID experiment will be installed in the same existing cryostat \cite{CUPID:2019imh}. Figure reproduced with permission from the CUPID collaboration.}
\end{figure}

The CUPID (``Cuore Upgrade with Particle IDentification") experiment \cite{CUPID:2019imh,CUPID:2022jlk} is being developed to search for \ndbd in the isotope $^{100}$Mo with $Q_{\beta\beta}=3034$ keV, which is above most naturally occurring $\gamma$ backgrounds. CUPID is based on cryogenic bolometer technology. The 240 kg of isotope $^{100}$Mo will be deployed in scintillating Li$_2$MoO$_4$ crystals, which convert emitted energy both into heat and light. This will allow for particle identification and thus the discrimination of background $\alpha$
from $\beta$ events, significantly reducing unwanted background. 
CUPID is the successor experiment to CUORE, a $^{130}$Te-based cryogenic bolometer featuring ca. 200 kg of isotope mass in the form of TeO$_2$ crystals, located at LNGS. CUPID will be installed in CUORE's cryostat. 
An engineering rendering of the cryostat with the current CUORE configuration is shown in Figure \ref{fig:cuore}.

Each of CUPID's Li$_2$MoO$_4$ crystals will be instrumented with  neutron transmutation doped germanium (NTD-Ge) thermal sensors. 
Together with NTD-based light detectors and Neganov-Trofimov-Luke amplification, sufficient pile-up rejection of coincident $2\nu\beta\beta$ events will be achieved.
See \cite{CUPID:2025avs} for a detailed discussion of the CUPID detector.  

CUPID has a projected $3\sigma$ discovery sensitivity of $1 \times 10^{27}$ yr corresponding to $\langle m_{\beta\beta}\rangle < 12-21$ meV \cite{CUPID:2025avs}. Beyond that, CUPID could be scaled up to CUPID-1T \cite{CUPID:2022wpt} containing a tonne of the isotope $^{100}$Mo with a projected $3\sigma$ discovery sensitivity of $9.1\times 10^{27}$ yr. This would require more than 10,000 channels to be read out, with multiplexed systems playing a crucial role in the readout solution.

While NTD-Ge thermal sensors meet CUPID's requirements, R\&D efforts are ongoing towards future upgrades of the detector or the deployment in CUPID-1T, in particular of faster light detectors with risetimes on the order of $\approx 100$ \textmu s. 
A promising candidate are Ir-Pt bilayer transition-edge sensors
(TESs), that are being developed at Berkeley and Argonne National Laboratories in the USA \cite{Singh:2022rck}. 

To minimize the heat load and maintain high radiopurity while scaling up the experiment with more detectors, CUPID will require less wiring per detector, which can be achieved through multiplexing techniques. In 2022, the Berkeley team asked the Cosmology Instrumentation Laboratory at McGill University to develop and provide the readout electronics for the CUPID TES effort. The McGill group specializes in digital frequency multiplexing (DfMUX) readout of superconducting detectors 
for mm-wave astronomy \cite{Bender:2014nnc}. Its custom FPGA-based ICE readout system \cite{Bandura:2016dpm} is deployed at multiple cosmology experiments all over the world, including the SPT-3G experiment at the South Pole in Antarctica. There, McGill's electronics are used to read out $\sim$16,000 TES detectors 
\cite{SPT-3G:2014dbx, SPT-3G:2021vps}, scanning the cosmic microwave background.

The same ICE electronics are planned to read out TES detectors for CUPID, and new firmware has been developed at McGill to meet the needs of the CUPID detectors, taking the SPT firmware \cite{Bender:2014nnc} as a starting point. Namely, since the CUPID TESs have to be read out much faster compared to SPT, the sampling rate has been increased by roughly 3 orders of magnitude, while the multiplexing ratio was lowered from 128x to 10--15x. Eventually, with 8 modules (i.e. 120 detectors) per board, 15 McGill ICE boards would be needed for the CUPID experiment, provided the TESs are selected as the detector technology for the scintillation light sensors in a future phase of CUPID.

%

\section{Conclusion}
The search for neutrinoless double beta decay is a global priority in the field of astroparticle physics. Current experiments reach half-life sensitivities of few $10^{26}$ years. Experiments with a significantly increased sensitivity are on the horizon with SNO+ and next-generation experiments CUPID, LEGEND-1000 and nEXO. SNOLAB, the Canadian low-background underground research facility, located 2\,km below the Canadian shield in Sudbury, Ontario, hosts SNO+ and is the preferred location for the nEXO experiment. Canadian researchers have been involved in searches for $0\nu\beta\beta$ for several decades. They play leading roles within the SNO+ and nEXO collaborations with the goal of making another breakthrough discovery in the neutrino sector. 
\section{Acknowledgements}
The authors thank Michel Adami\v{c}, Erica Caden, Mark Chen, Matt Dobbs, Giorgio Gratta, Aksel Hallin, Karsten Heeger, Pillalamarri Jagam, Chris Jillings,  Ryan Martin, Art McDonald, Maura Pavan, David Sinclair, Fran Spidle, and Simon Viel for valuable input and discussions.

Competing interests: The authors declare there are no competing interests. The data used to generate Figure\,\ref{fig:lobster} is available upon request to Alexander Wright.
\bibliography{CJP6_0vBB_Decay}
\end{document}